\documentclass[reprint,twocolumn,superscriptaddress,secnumarabic,amssymb, nobibnotes, aps, prb]{revtex4-2}

\usepackage{amsmath}    % need for subequations
\usepackage{graphicx}    % need for figures
\usepackage{verbatim}    % useful for program listings
\usepackage{color}           % use if color is used in text
\usepackage{subfigure}   % use for side-by-side figures
\usepackage{hyperref}    % use for hypertext links, including those to external documents and URLs
\usepackage{mathrsfs}
\usepackage{braket}
\usepackage{physics}
\usepackage{graphicx}    % need for figures
\usepackage{xspace}

\newcommand{\opd}{\operatorname{d}}
\newcommand{\pd}{\phantom\dagger}
\newcommand{\bR}{\mathbf{R}}
\newcommand{\br}{\mathbf{r}}

\newcommand{\bq}{\mathbf{q}}

\newcommand{\jh}{J_{\rm H}}
\def\sro{SrRu$_{2}$O$_{6}$\xspace}

\begin{document}

\title{Hidden covalent insulator and spin excitations in \sro}%

\author{Diana Csontosov\'a}
\affiliation{Department of Condensed Matter Physics, Faculty of
  Science, Masaryk University, Kotl\'a\v{r}sk\'a 2, 611 37 Brno,
  Czechia}
\affiliation{Institute for Solid State Physics, TU Wien, 1040 Vienna, Austria}
\author{Ji\v{r}\'{\i} Chaloupka}
\affiliation{Department of Condensed Matter Physics, Faculty of
  Science, Masaryk University, Kotl\'a\v{r}sk\'a 2, 611 37 Brno,
  Czechia}
\author{Hiroshi Shinaoka}
\affiliation{Department of Physics, Saitama University, Saitama 338-8570, Japan}
\author{Atsushi Hariki}
\affiliation{Department of Physics and Electronics, Graduate School of Engineering, Osaka Metropolitan University, 1-1 Gakuen-cho, Nakaku, Sakai, Osaka 599-8531, Japan}
\author{Jan Kune\v{s} }
\affiliation{Department of Condensed Matter Physics, Faculty of
  Science, Masaryk University, Kotl\'a\v{r}sk\'a 2, 611 37 Brno,
  Czechia}
  \affiliation{Institute for Solid State Physics, TU Wien, 1040 Vienna, Austria}

\date{\today}%a

\begin{abstract}
The density functional plus dynamical mean-field theory is used to study the spin excitation spectra of \sro. A good quantitative agreement with experimental spin excitation spectra is found. 
Depending on the size of the Hund's coupling $\jh$
the systems chooses either Mott insulator or covalent insulator state when magnetic ordering is not allowed. We find that the nature of the paramagnetic state has negligible influence on the charge and spin excitation spectra. We find that antiferromagnetic correlations hide
the covalent insulator state for realistic choices of the interaction
parameters.
\end{abstract}

\maketitle
%\tableofcontents

%%%%%%%%%%%%%%%%%%%%%%%%%%%%%%%%%%%%%%%%%%%%%%%%%%%%%%%%%%%%%%%%%%%
%%%%%%%%%%%%%%%%%%%%%%%%%%%%%%%%%%%%%%%%%%%%%%%%%%%%%%%%%%%%%%%%%%%
\section{Introduction}
Competition between 
%inter-atomic hopping and intra-atomic electron-electron interaction 
kinetic and interaction energy 
is the corner stone of the correlated electrons physics.
In the paradigmatic bandwidth control scenario of Hubbard model at half filling,
increasing the interaction--to--bandwidth ratio suppresses the charge fluctuations
and eventually drives the system to a Mott insulator (MI) state~\cite{Brinkmann1970}.
Real materials provide variations on this theme~\cite{Pavarini2004,Kim2008}, but 
also alternative mechanisms of correlation driven metal-insulator transition (MIT) 
such as site-selective Mott transition~\cite{Park2012},
spin-state crossover~\cite{Patterson2004,Kunes2008a}, Kondo insulator~\cite{Fisk1996},
or gapping the ligand bands~\cite{Kunes2010}
to name a few. Often the paramagnetic (PM) MIT
is hidden by a magnetic long-range order, which raises
the question how much about the nature of the PM phase can be learned from the properties of
the ordered phase. The studies of single-band Hubbard model~\cite{Sangiovanni2006,Frantino2017}
found rather subtle differences in anti-ferromagnetic (AFM) phase on the two sides of the Mott transition,
which can be difficult or even impossible to identify in multi-orbital setting of real materials. 

A weakly correlated state does not have to be metallic in order to exhibit charge fluctuations. A covalent insulator (CI)~\cite{Fu1995,*Kunes2008,*Sentef2009}, with a gap between bonding and anti--bonding states does as well. 
Mazin~{\it et al.}~\cite{Mazin2012} pointed out 
a special hopping pattern of $t_{2g}$ electrons in layered transition metal oxides with honeycomb lattice and edge-sharing octahedra such as Na$_2$IrO$_3$, $\alpha$-RuCl$_3$, Li$_2$RuO$_3$ or \sro.
Considering only the dominant hopping paths between the nearest-neighbor
metal ions, the $t_{2g}$ electrons are trapped on the hexagonal structural units, 
which gives rise to molecular orbitals clearly visible in the 
calculated non-interacting electronic spectra.
At half filling the Fermi level falls into the band gap
between the molecular peaks~\cite{Streltsov2015}, which stabilizes the CI state.
On the other hand, the tendency to form a high-spin MI is maximal also at half filling~\cite{georges2013}, which leads to a competition without an {\it a priori} winner.

\begin{figure}
%\begin{center}
 \includegraphics[width=0.8\columnwidth]{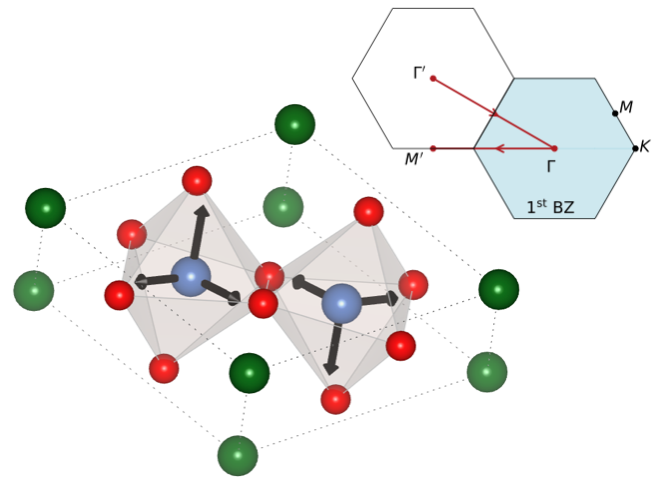}
%\end{center}
\vspace{-0.4cm}
\caption{The unit cell of \sro: Ru (blue), O (red) and Sr (green) atoms,
 visualized using VESTA3 \cite{vesta}. The arrows mark the local  orbital coordinates. Path in the reciprocal space used for plotting magnon dispersions.}
\label{fig:crys}
\end{figure}
This scenario is realized in \sro with nominally $t_{2g}^3$ configuration.
An antiferromagnetic insulator with high N\'eel temperature $T_{\rm N}$ of 563~K~\cite{Hiley2014},
it does not exhibit the Curie-Weiss susceptibility in the PM phase.
Instead, the susceptibility increases up to the highest reported temperature of about
730~K~\cite{Hiley2015}. Classification of \sro based on numerical studies has been controversial.
Streltsov~{\it at al.}~\cite{Streltsov2015} performed 
density functional plus dynamical mean-field theory
(DFT+DMFT) calculations for Hund's coupling
$\jh =0.3$~eV. Pointing out the discrepancy between the theoretical ionic moment of 3~$\mu_{\rm B}$, 
a value essentially reproduced by their DFT+DMFT, 
and the observed ordered moment of 1.4~$\mu_{\rm B}$
they argued that the electronic structure of \sro is dominated by molecular orbitals. 
Hariki~{\it et al.}~\cite{Hariki2017} using a similar DFT+DMFT approach found a crossover
between CI and MI in the PM phase for $\jh$ between $0.16-0.19$~eV, depending on temperature. 
They also found that in the AFM phase the size of the ordered moment is essentially the same 
on both sides of the CI/MI crossover and agrees well with experimental as well as the DFT
value, when the overlaps of Wannier orbitals are properly accounted for. 
The uncertainty in the value of the Hund's exchange $\jh$ thus left the question of electronic structure of \sro open.

Using resonant inelastic x-ray scattering (RIXS) to
map out the magnon dispersion Suzuki~{\it et al.}~\cite{Suzuki2019} 
concluded that \sro is a Mott insulator 
because the magnon spectrum can be well described by $S=3/2$ Heisenberg model with parameters 
obtained by strong-coupling expansion with first principles hopping parameters. 
They pointed out the difference between a large paramagnetic N\'eel temperature $\Theta$, proportional to the 
inter-atomic exchange $J$ and reflected in the magnon bandwidth, and the smaller ordering temperature $T_N$,
determined by the spin gap. They argued that the observed absence of Curie-Weiss behavior above $T_N$ is consistent
with the behavior of 2D Heisenberg model, for which it is expected first for $T>\Theta$.

We compute the spin excitation spectra~\cite{Park2011,Boehnke2013} using DFT+DMFT~\cite{Kotliar2006}. We pursue two objectives
(i) apply the DMFT approach to dynamical susceptibilities based of Bethe-Salpeter equation (BSE)~\cite{Kunes2011,Niyazi2021} to an ordered state of a real material and assess its quantitative accuracy, 
(ii) analyze the connection between the character of the PM phase, MI vs CI, and the properties of the AFM phase.
The DMFT BSE approach has been successfully applied to antiferromagnetic magnons in up to 3-orbital model~\cite{Niyazi2021}. Here we focus on quantitative comparison with experiment, the role of spin-orbit coupling (SOC), the relationship between single-ion anisotropy and the spin gap, and other spin excitations beyond magnon. In order to address (ii), we vary $\jh$ across the CI--MI crossover.

\section{Computational method}
We study the `$t_{2g}$-only' model of Ref.~\onlinecite{Hariki2017} with Slater-Kanamori interaction obtained by wannierization~\cite{wannier90,wien2wannier} from density functional calculation~\cite{wien2k}.
Unlike in Ref.~\onlinecite{Hariki2017} we use the basis of $xy$, $yz$ and $xz$ Wannier orbitals in the coordinates shown in Fig.~\ref{fig:crys}, see Supplemental Material (SM)~\cite{sm} 
and references~\cite{Georges1996,gull11,Stavropoulos2021,JarrellGubernatis96,Perdew96,wang09,LiuHYK2022} therein. In order to reduce the computational effort, the calculations were done for C-type (2 atoms) rather than the experimental G-type (4 atoms) structure. This approach is justified by the miniscule inter-layer coupling~\cite{Hiley2015}.

Several Ru compounds with honeycomb structure exhibit Ru-Ru dimerization 
upon cooling,  Li$_2$RuO$_3$~\cite{Miura2007}, or under pressure,
RuCl$_3$~\cite{Bastien2018} or Ag$_3$LiRu$_2$O$_6$~\cite{Takayama2022}, which has been associated 
with a delicate balance between covalent bonding and spin-orbit entanglement~\cite{Takayama2022}. Such behavior was observed neither in \sro~\cite{Hiley2015} 
nor in isoelectronic BaRu$_2$O$_6$~\cite{Marchandier2020}. We speculate that this is related to the Ru $d^3$ configuration, which favors high-spin moment over spin-orbit entanglement (relative to other filling) and corresponds to one electron per each Ru-Ru bond. Therefore we perform our calculations for fixed experimental structure.

Throughout this study we keep the interaction parameter $U=2.7$~eV fixed and vary 
$\jh=0.16-0.22$~eV as well as temperature. In PM calculation we enforce the spin symmetry
of the self-energy in each DMFT iteration. 

The DMFT~\cite{Metzner1989,*Georges1992,*Jarrell1992} calculations were performed with
a multiorbital implementation~\cite{SHINAOKA2017} of the continuous-time hybridization expansion Monte Carlo method~\cite{Werner2006a} based on ALPS core libraries~\cite{Gaenko2017, *Bauer2011}. Some of the DMFT calculations were benchmarked
against results obtained with DCore~\cite{Shinaoka2021}. The BSE with local particle-hole irreducible
vertex~\cite{Zlatic1990} was solved for the lowest 10 bosonic Matsubara frequencies in the Legendre representation~\cite{Boehnke2011}.
%with up to $N = 40$ Legendre coefficients to
%obtain the 2P correlation function $X_{ijs,kls'}(\bq,i\omega_n)$~\cite{Niyazi2021,sm}.
%Here, $i$, $j$, $k$, $l$ are the spin/orbital indices, $s$ and $s'$ the sublattice
%indices. 
The desired dynamical susceptibilities
$\langle O_{-\mathbf{q}} O_{\mathbf{q}} \rangle_{\omega}$ 
were obtained by sandwiching the general 2-particle susceptibility
with the corresponding vertices followed by analytic continuation~\cite{geffroy2019, LEVY2017}
%~\textcolor{red}{REFERENCE}
, see SM~\cite{sm} for details.
The reciprocal space operators are related to local observable by
the Fourier transform
\begin{equation}
    O_{\bq} = \sum_{\bR,s} e^{-i\bq\cdot(\bR+\br_s)} O_{\bR s}\quad \br_s=
    \begin{cases}
    (\tfrac{2}{3},\tfrac{1}{3},0)\ s\!=\!\rm{A}\\
    (\tfrac{1}{3},\tfrac{2}{3},0)\ s\!=\!\rm{B},
    \end{cases}
\end{equation}
where the index $s$ refers to the two Ru sites in the unit cell. In the following
we study the transverse spin susceptibility with $O\equiv S^x$, and $S=3/2\rightarrow 1/2$
excitations, for which we choose a representative operator $O\equiv X$ below, generating $\Delta S^z =\pm 1$ transitions between $S=3/2$ and $S=1/2$ manifolds
\begin{align}
    S^x_{\bR s}& = \sum_{\alpha=1}^3 \opd^\dagger_{\bR s \alpha\uparrow}\opd^{\pd}_{\bR s\alpha\downarrow}+H.c.\label{eq:Sx}\\
    X_{\bR s}& = \left(\opd^\dagger_{\bR s 1\uparrow}\opd^{\pd}_{\bR s 1\downarrow}-
    \opd^\dagger_{\bR s 2\uparrow}\opd^{\pd}_{\bR s 2\downarrow}\right)+H.c.\label{eq:X}
\end{align}
The operator $X$ is chosen to be representative of a set of closely spaced transitions, see SM~\cite{sm}.
%%%%%%%%%%%%%%%%%%%%%%%%%%%%%%%%%%%%%%%%%%
\begin{figure}%\vspace{+0.5cm}
\begin{center}
  \includegraphics[width=\columnwidth]{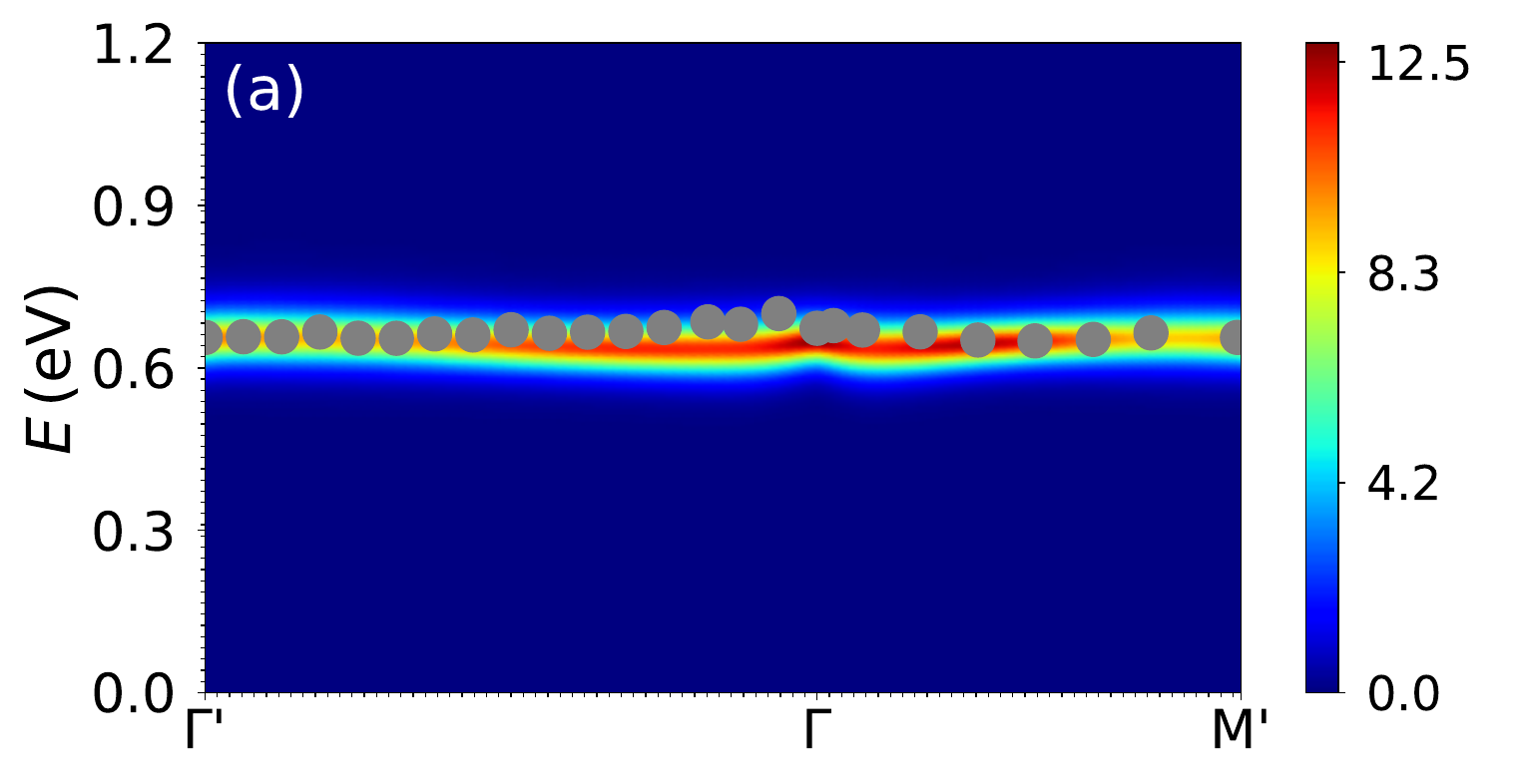}
  \includegraphics[width=\columnwidth]{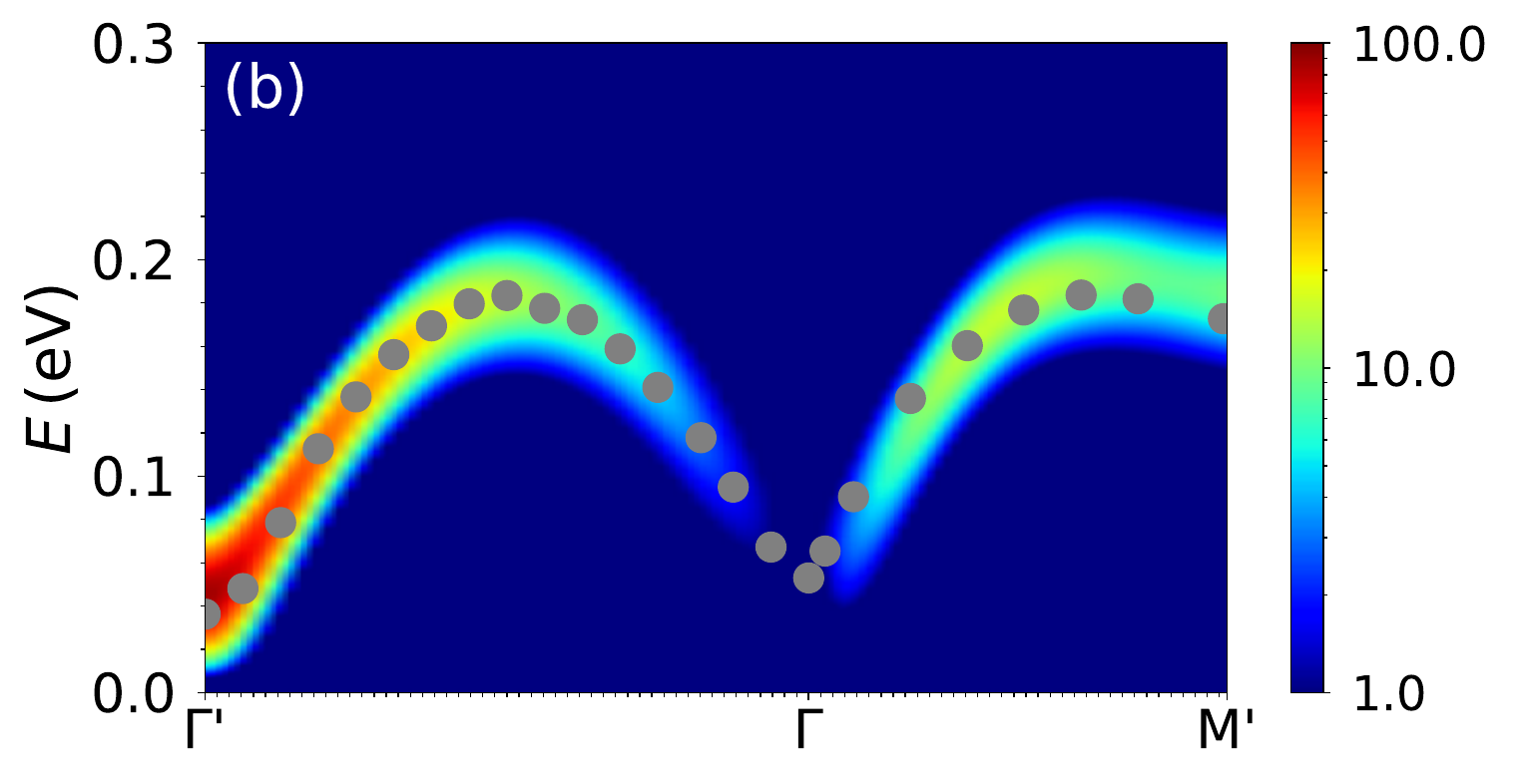}
\end{center}
\vspace{-0.5cm}
\caption{Imaginary part of the dynamical susceptibility along the $\Gamma^{\prime}$ - $\Gamma$ - $M$ path shown in Fig.~\ref{fig:crys}
for $\jh=0.16$~eV at $T=464$~K. Grey dots denote the maxima
of the corresponding RIXS features~\cite{Suzuki2019}. Top (linear color scale): $\langle X_{-\mathbf{q}} X_{\mathbf{q}} \rangle_{\omega}$ representing the $S=3/2\rightarrow 1/2$ transitions.
Bottom (logarithmic color scale): $\langle S^x_{-\mathbf{q}} S^x_{\mathbf{q}} \rangle_{\omega}$ 
corresponding to magnon.}
%\vspace{-0.4cm}
\label{fig:theo_vs_exp}
\end{figure}
%%%%%%%%%%%%%%%%%%%%%%%%%%%%%%%%%%%%%%%%%%
\begin{figure}%\vspace{+0.5cm}
\begin{center}
   \includegraphics[width=\columnwidth]{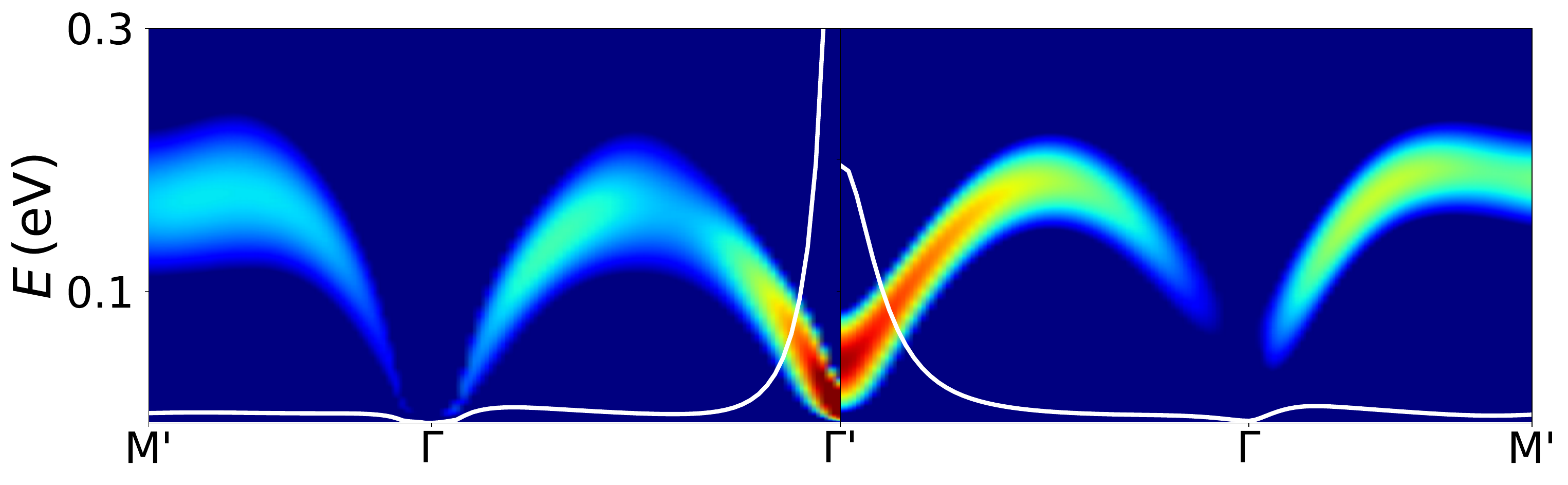}
\end{center}
\vspace{-0.5cm}
\caption{Effect of the spin-orbit coupling (SOC) on the
magnon spectra:  with SOC (right panel) and without SOC (left panel). 
Color scale is the same as in Fig.~\ref{fig:theo_vs_exp}b)
The white line is a spectral weight computed as $\Omega_{\mathbf{q}} = -\tfrac{1}{\pi}\int_0^{0.3} \mathrm{d}\omega 
\operatorname{Im}\langle S^x_{-\mathbf{q}} S^x_{\mathbf{q}} \rangle_{\omega}$.
The results were obtained 
for $\jh  = 0.16\,\mathrm{eV}$ and $T = 464\,\mathrm{K}$.} 
\vspace{-0.4cm}
\label{fig:w_vs_w/o_soc}
\end{figure}

%ADD CODES wien2k, wannier90, w2w, ALPS, dcore?, link to our BSE solver

\section{Results and discussion}
\subsection{Magnon dispersion}
The DMFT calculations lead to AFM with out of plane orientation of the local moment for temperatures
below 1500~K. Since the magnetism of \sro is essentially 2D~\cite{Hiley2015,Suzuki2019} this overestimation
by DMFT is expected. The DMFT does not obey the Mermin-Wagner theorem and the calculated ordering temperature represents
$\Theta$ rather than $T_N$. This does not mean that the DMFT AFM solution should not be able to capture the 
ordered state of the real material.
Fig.~\ref{fig:theo_vs_exp} shows a comparison of the dynamical suspcetibilities 
$\langle X_{-\mathbf{q}} X_{\mathbf{q}} \rangle_{\omega}$ and 
$\langle S^x_{-\mathbf{q}} S^x_{\mathbf{q}} \rangle_{\omega}$ calculated in the AFM phase
at 464~K to the experimental RIXS data~\cite{Suzuki2019}. 
The magnetic moments at this temperature are essentially saturated~\cite{Hariki2017,sm} and thus no significant change in the computed spectra is expected upon further cooling.
Rather than computing the full RIXS spectra, calculation of which would require evaluation
of transition amplitudes~\cite{Hariki2020,Haverkort10} with the possibility
of multi-particle excitations~\cite{Nag2020,Li2023} and is not possible with the present methods, we compare the dispersions of specific spectral features. 
We find a very good match of the magnon dispersion including
the bandwidth, the spin gap and the distribution of spectral weight. The magnon bandwidth
of $183$~meV corresponds to the effective nearest-neighbor exchange $JS=61\,\mathrm{meV}$ between $S=3/2$ local moments.
%
%{\color{blue} JIRI}
%

A straightforward strong-coupling calculation with the same parameter setup
yields a remarkably similar value $JS\approx 66\:\mathrm{meV}$ \cite{sm},
essentially unaffected by SOC. However, by inspecting the exact solution of
our Hubbard model on a single bond \cite{sm}, we found the spin $S=3/2$ picture
to be significantly disturbed by a large involvement of higher multiplet
states at energies $\gtrsim 3J_\mathrm{H}$ \cite{sm}. In such situation, the
DMFT approach covering the entire spectrum of multiplet states is highly
advantageous.

The spin gap of approximately 45~meV is related to the single-ion anisotropy
$\Delta_{\text{SIA}}=E_{\pm 1/2}-E_{\pm 3/2}=6.6\:\mathrm{meV}$, defined as
the difference between the atomic states belonging to the $S=3/2$
multiplet~\footnote{The spin quantization axis is pointing out of plane}. 
The strong-coupling evaluation of SIA suggests that the above ionic value is
actually strongly renormalized by exchange processes \cite{sm}.
Within the linear spin-wave theory of Heisenberg antiferromagnet, the large
gap is easily explained even for small SIA, as it is given by
$S\sqrt{6J\Delta_\mathrm{SIA}}$~\cite{Suzuki2019}. 
Nevertheless, it is not self-evident that the present numerical approach must
capture it accurately. 
%
%The spin gap of approximately 45~meV is substantially enhanced compared to the single-ion anisotropy %$\Delta_{\text{SIA}}=E_{\pm 1/2}-E_{\pm 3/2}$= 6.6~meV, defined as the difference between the atomic states %belonging to the $S=3/2$ multiplet~\footnote{The spin quantization axis is pointing out of plane}. This enhancement %is understood within the linear
%spin-wave theory of Heisenberg antiferromagnet~\cite{Suzuki2019}. Nevertheless, it is not self-evident that the %present numerical approach must capture it accurately. 
%
%%%%%%%%%%%%%%%%%%%%%%%%%%%%%%%%%%%%%%%%%%%%%%%%%%%%%%%%%%%%%%%%%%%

%%%%%%%%%%%%%%%%%%%%%%%%%%%%%%%%%%%%%%%%%%
\begin{figure}
%\vspace{+0.5cm}
\begin{center}
   \includegraphics[width=0.9\columnwidth]{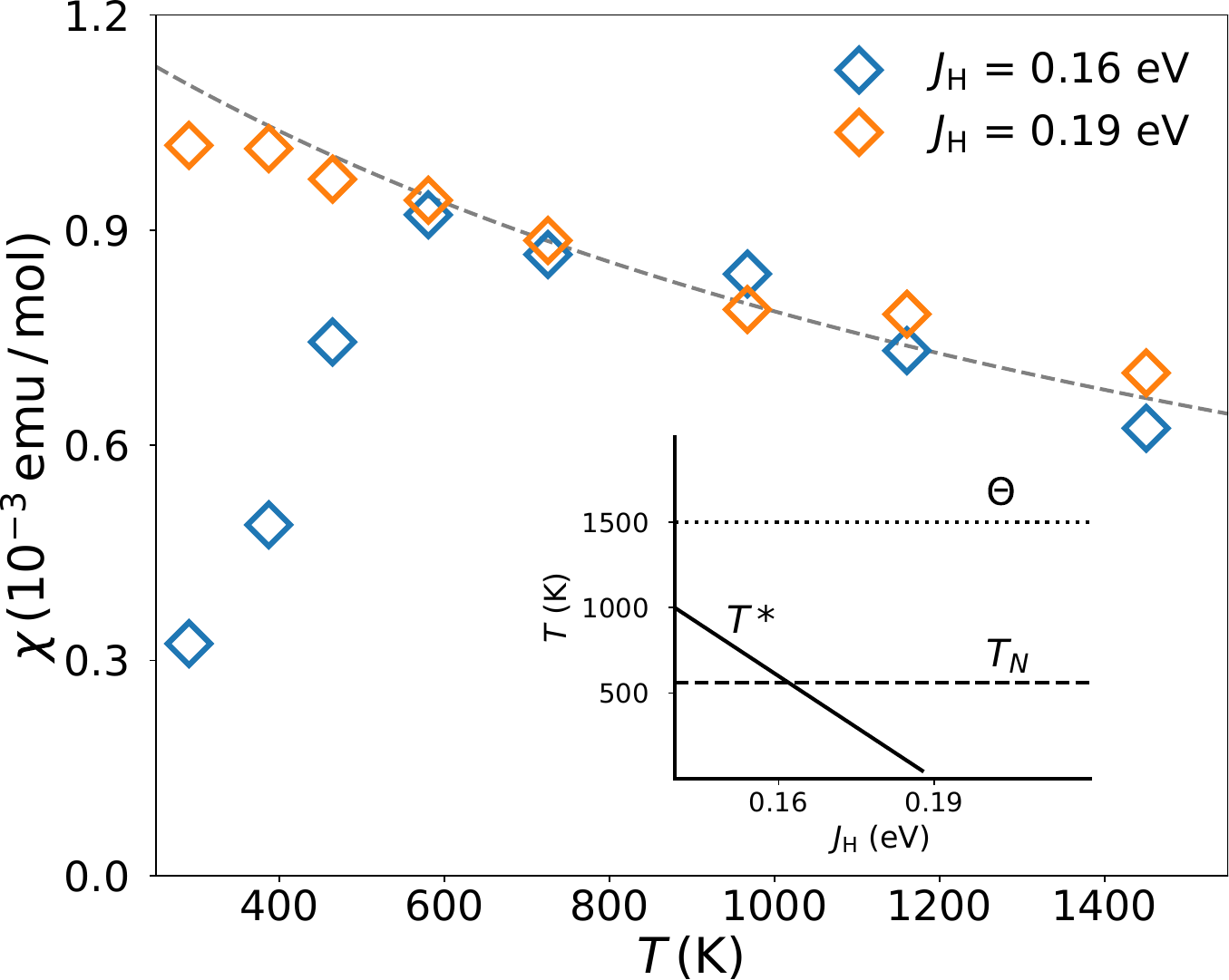}
   \includegraphics[width=0.9\columnwidth]{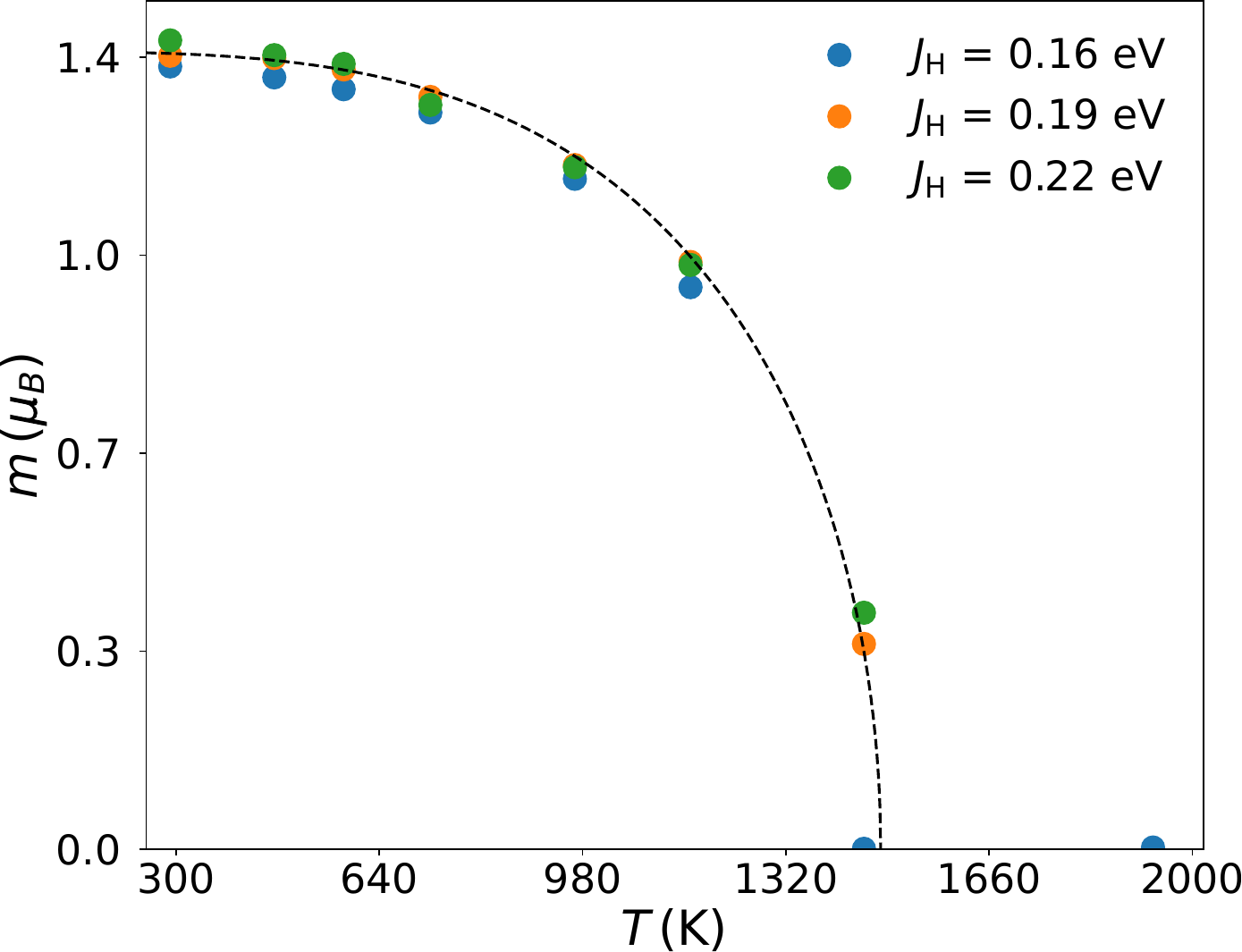}
\end{center}
\vspace{-0.5cm}
\caption{Top panel: Uniform susceptibility for $J_{\rm H} = 0.16\,\mathrm{eV}$ and $\jh = 0.19\,\mathrm{eV}$ in the PM state. 
%The vertical dashed line shows the experimentally determined critical temperature $T_{\rm N} = 563\,\mathrm{K}$~\cite{Hiley2015}. 
The dashed line shows the Currie-Weis susceptibility $\chi \propto (T + \Theta)^{-1}$ with $\Theta=1480\,\mathrm{K}$.  Magnitude of the calculated $\chi(T)$ is about 30\% smaller than the experimental one~\cite{Hiley2015}. Inset: a cartoon picture
of different temperature scales in \sro.
Bottom panel: Order moment as a function of temperature for the 
studied values of $\jh$.}
%}
\vspace{-0.4cm}
\label{fig:chi_stat}
\end{figure}
%%%%%%%%%%%%%%%%%%%%%%%%%%%%%%%%%%%%%%%%%%%%%%%%%%%%%%%%%%%%%%%%%%%
\begin{figure*}
\vspace{+0.5cm}
\begin{center}
   \includegraphics[width=\textwidth]{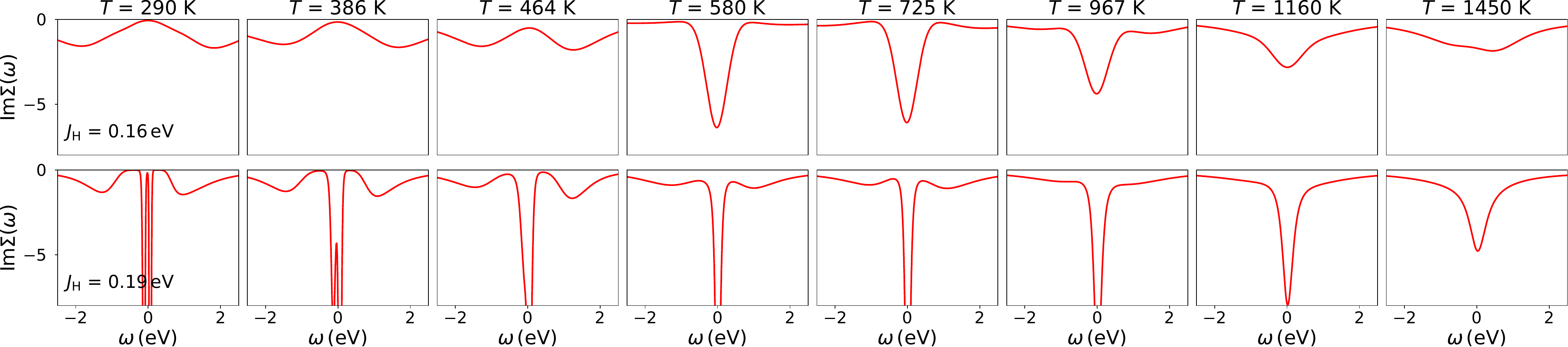}
\end{center}
\vspace{-0.5cm}
\caption{Imaginary part of self-energy $\mathrm{Im}\Sigma_{ii} (\omega)$ 
(diagonal element) on the real frequency axis for $\jh = 0.16\,\mathrm{eV}$ (top row) and $\jh = 0.19\,\mathrm{eV}$ (bottom row) at various temperatures $T$. The crossover temperature $T^\star$ for $\jh$=0.16~eV lies between 464~K and 580~K.}
\label{fig:selfEneryReal}
\end{figure*}
\begin{figure}
\vspace{+0.5cm}
%\begin{center}
   \includegraphics[width=\columnwidth]{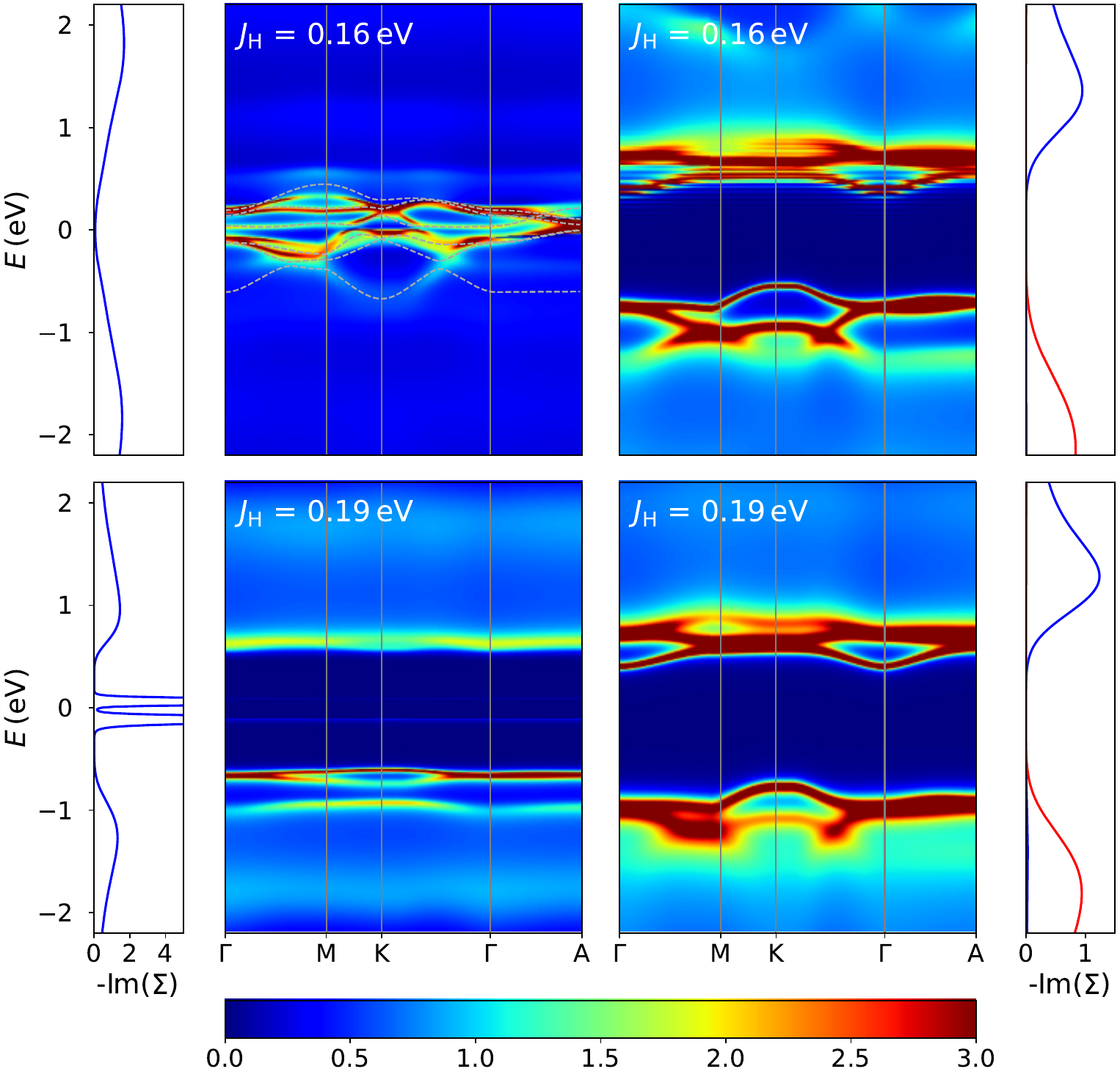}
%\end{center}
\vspace{-0.5cm}
\caption{Spectral functions and corresponding imaginary parts of self-energies on the real axis in the constrained PM solution (left half) and AFM solution (right half). The calculations were performed for $T = 290\,\mathrm{K}$. Red and blue color in the figures with AFM self-energies distinguish between spin up and spin down component. The grey lines in the spectral function with $\jh = 0.16\,\mathrm{eV}$ show DFT band structure squeezed by factor 2.2. } 
\vspace{-0.4cm}
\label{fig:1Pspectra}
\end{figure}
\begin{figure*}[t]%\vspace{+0.5cm}
\begin{center}
   \includegraphics[width=0.9\textwidth]{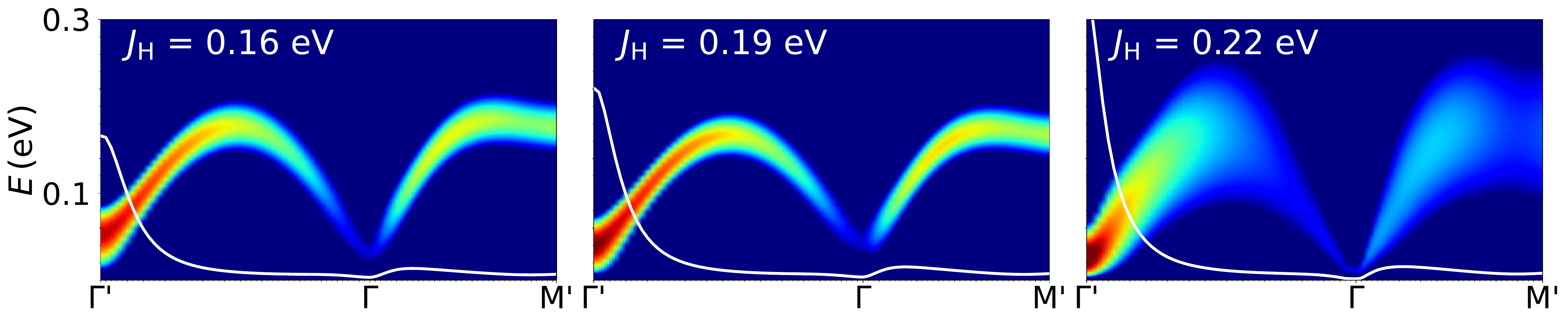}
\end{center}
\vspace{-0.5cm}
\caption{Comparison of magnon spectra in covalent insulator ($J_{\mathrm{H}}  = 0.16\,\mathrm{eV}$) and Mott insulator ($J_{\mathrm{H}}  = 0.19\,\mathrm{eV}$ and $J_{\mathrm{H}}  = 0.22\,\mathrm{eV}$) phases. The calculations were performed for $T = 464\,\mathrm{K}$. The spin gaps for $J_{\mathrm{H}}  = \{0.16, 0.19, 0.22\}\,\mathrm{eV}$ are $\Delta_{\mathrm{m}}  = \{45, 36, 35\}\,\mathrm{meV}$, respectively. The white line is a spectral weight $\Omega_{\mathbf{q}}$.(The same color scale as Fig.~\ref{fig:theo_vs_exp}b)} 
\vspace{-0.4cm}
\label{fig:Js}
\end{figure*}
%%%%%%%%%%%%%%%%%%%%%%%%%%%%%%%%%%%%%%%%%%%%%%%%%%%%%%%%%%%%%%%%%%%
We have also carefully checked the 
out-of-plane orientation of the ordered moments, see SM~\cite{sm}, and verified its origin in SOC by
performing calculations with SU(2)-symmetric Hamiltonian without SOC. As expected
we find two gapless linear Goldstone modes with divergent spectral weights in this case, see Fig.~\ref{fig:w_vs_w/o_soc}. 
%To compare the amplitudes of Goldstone mode's for caclulations performed without and with SOC, we plot the spectral weithts $\Omega_{\mathbf{q}} = -\tfrac{1}{\pi}\int_0^{0.3} \mathrm{d}\omega 
%\operatorname{Im}\langle S^x_{-\mathbf{q}} S^x_{\mathbf{q}} \rangle_{\omega}$.

The experimental RIXS spectra~\cite{Suzuki2019} exhibit a prominent low-energy 
feature associated with $S=3/2\rightarrow 1/2$ transitions. Our calculations, Fig.~\ref{fig:theo_vs_exp}, reproduce the position of this feature fairly well, although the SOC induced mixing with the low energy magnon limits the resolution of 
the higher energy structures.

%While we do not attempt
%to calculate the complete RIXS spectra including the transition amplitudes, we evaluate
%the correlation functin of the operator (\ref{eq:X}) associated with this transition in
%. 

%Having established the relevance of our results for \sro we proceed to discuss their physical consequences. The authors of Ref.~\onlinecite{Suzuki2019} argued that \sro is a Mott insulator
%because the S=3/2 Heisenberg model with nearest neighbor interactions properly captures the
%spectra of AFM phase. We will demonstrate that this implication does not necessarily apply.
%The observed behavior of the AFM phase does not allow to draw conclusions about the nature of the PM phase.

%%%%%%%%%%%%%%%%%%%%%%%%%%%%%%%%%%%%%%%%%%%%%%%%%%%%%%%%%%%%%%%%%%%

\subsection{Mott vs covalent insulator}
In calculations performed in the PM state,
the authors of Ref.~\onlinecite{Hariki2017} 
observed a crossover between the low-temperature CI and high temperature MI
at a scale $T^\star$, which strongly depends on $\jh$. 
For $\jh= 0.16$~eV the scale $T^\star$ lies in the 600--800~K range, while for
$\jh\gtrsim 0.19$~eV only MI was observed.
\sro exists in the PM phase below 800~K, however, since DMFT exaggerates  
its ordering temperature~\footnote{As discussed in Ref.~\onlinecite{Suzuki2019} \sro
is essentially a 2D material with $T_{\rm N}$ determined by the spin gap. The dimensionality
aspect is not properly captured by DMFT.}.
we enforce the PM solution by constraint, in order to study it at lower temperatures.

The different temperature scales discussed below are summarized in the inset of Fig.~\ref{fig:chi_stat}. The paramagnetic N\'eel temperature $\Theta$, 
which we identify with the DMFT ordering temperature, is estimated from the present study, the bottom panel of Fig.~\ref{fig:chi_stat}, and Ref.~\onlinecite{Hariki2017}. The CI/MI crossover temperature
$T^\star$ is estimated from Ref.~\onlinecite{Hariki2017} and the uniform susceptibility
from $\jh= 0.16$~eV of this study. Finally, $T_N$ is the experimental ordering temperature,
the weak $\jh$-dependence of which may deduced from the behavior of the spin gap as a function of $\jh$.

Next, we discuss the properties of the constraint PM solutions.
At high temperatures ($T>T^\star$) CI and MI behave similarly.
The imaginary part of the self-energy, shown in Fig.~\ref{fig:selfEneryReal}, exhibits a broad peak at the chemical potential, which give rise to a gap containing some incoherent spectral weight.
At low temperatures ($T<T^\star$) CI and MI are distinguished by several characteristics. 
The self-energy of CI has a Fermi liquid character with vanishing imaginary part at the chemical potential. The peak in MI self-energy 
becomes sharper and its background vanishes in the low-energy region, 
which defines the Mott gap. This gives rise to distinct band structures shown in Fig.~\ref{fig:1Pspectra}. For the  evolution of the self-energy on the imaginary axis see SM~\cite{sm}. 
The CI and MI respond differently to a magnetic field.
The magnetic susceptibility $\chi(T)$ of MI, in Fig.~\ref{fig:chi_stat},  exhibits
the usual Curie--Weiss decrease with increasing temperature. The high-temperature
susceptibility of CI follows the same trend. However, once the Fermi liquid behavior sets in below $T^\star$~\footnote{Note that calculations in Ref.~\onlinecite{Hariki2017} put $T^\star$ to around 800~K.} the susceptibility starts to drop, which gives rise to a broad maximum. A positive slope of the experimental $\chi(T)$ above the transition temperature was pointed out by the authors of Ref.~\onlinecite{Hiley2015}.
The CI and MI states are also distinguished by local charge fluctuations on the Ru site~\cite{Hariki2017}. This is reminiscent of the site selective
Mott transition~\cite{Park2012}, where both CI- and MI-like sites are found within
the same compound. Numerical simulations of core-level spectroscopies such as x-ray absorption or RIXS~\cite{Green2016,*Lu2018,*Winder2020} revealed distinct dependencies on the  incoming photon frequency. Similar spectroscopic signature may expected for
the CI and MI states.

How is the different character of the PM phase reflected in the AFM phase?
Upon magnetic ordering the self-energy is dominated by
the spin-dependent Hartree shift and electronic spectra for large and small $\jh$ in Fig~\ref{fig:1Pspectra}
resemble one another. In Fig.~\ref{fig:Js} we compare the magnon spectra obtained at 464~K for $\jh$ values on both sides of CI/MI crossover. A difference is hardly noticeable. 
%Varying width of the spectral lines is attributed to an uncertainty of the analytic continuation rather than physics of the system. 
There is a discernible trend of decreasing spin gap with $\jh$, which follows from the behavior of the single-ion anisotropy. Overall the parameters extracted using strong-coupling theory describe the
magnons equally well on CI and MI side in the parameter space.
%We point out that a similar good match between DMFT and the strong-coupling limit 
%for the excitation dispersion in the ordered phase was observed in Ref.~\onlinecite{Geffroy2019a}.

Can the behavior of the CI susceptibility explain the experimentally observed behavior of $\chi(T)$ in the PM phase? 
%Can one argue that the increase of $\chi(T)$ in the CI parameter range is obscured by the overestimated transition temperature $T_N$ and if 
Is it plausible that an improved theory, which pushes the calculated $T_N$ to its experimental value below $T^\star$, uncovers the CI susceptibility?
We argue that it is not.
The problem of DMFT description is not quantitative overestimation of $T_N$ because of
inaccurate treatment of the 3D aspect (inter-layer coupling) of the material. In fact
the estimated inter-layer coupling~\cite{Hiley2015} was shown to be by far too small to account
for the observed $T_N$~\cite{Suzuki2019}. 
The problem is a conceptual inefficacy to distinguish between the paramagnetic N\'eel temperature $\Theta$ and the ordering temperature $T_N$. In fact
the $\Theta$ given by DFT+DMFT, i.e., the onset of strong AFM correlations, is likely correct as suggested by the correct magnon bandwidth obtained in the calculation. DMFT does not exaggerate
the onset temperature of the AFM correlations, but describes
them as static (AFM order), while in the 2D reality they remain dynamical
down to much lower temperature $T_N$ determined by a spin gap. 
Although the spin gap itself is well captured, its effect on $T_N$
is completely missing in the theory.
The CI physics can be realized if the crossover temperature $T^\star$
is above the onset of AFM correlations $\Theta$.
In the present case for smaller $\jh$ we get $T_N < T^\star<\Theta$ and thus
the increase of $\chi(T)$ above $T_N$ represents the physics of 2D Heisenberg magnet
rather than that of CI.

We would like to point out the analogy of the present physics with the
Kondo lattice model~\cite{Doniach1977}. In both cases a local moment disappears below a certain temperature, 
$T^\star$ in CI or Kondo temperature in case of the Kondo lattice, if not correlated to other moments on the lattice.
In both cases, inter-site correlations between the local moments can preclude their disappearance if sufficiently strong, which we conjecture to mean $T^\star < \Theta$ in the present case. These are examples of a situation when inter-site
interaction between the local excited states (carrying the local moments), eliminates the (non-magnetic) local ground states from the set of global low-energy states.

\section{Conclusions} 
 We have calculated the spin excitation spectra of \sro using 
 DFT+DMFT approach and found a quantitative match with the experimental observations~\cite{Suzuki2019}, notably for the spin gap due to the spin-orbit coupling. The paramagnetic state of \sro, depending on the strength of the Hund's coupling $\jh$, exhibits either covalent insulator or Mott insulator characteristics below $T^\star\approx 580$~K.
 Once in the AFM ordered state the magnon and electron excitation spectra 
 are essentially the same for $\jh$ on both sides 
 of the covalent insulator / Mott insulator crossover. 
 Our calculations for realistic $\jh$ on both sides of the CI/MI crossover
 lead to the conclusion that $T^\star$ is substantially below the
 temperature $\Theta$ at which the AFM correlations set in
 and therefore the covalent insulator state remains always 'hidden'.

\begin{acknowledgments}
The authors thank H. Suzuki for providing the experimental data of Fig.~\ref{fig:theo_vs_exp}, A. Kauch for critical reading of the manuscript, and K.-H. Ahn for valued discussions in the early stage of this work.
This work has received funding from
QUAST-FOR5249 project I 5868-N (D.C., J.K.) of the Austrian Science Fund (FWF),
Czech Science Foundation (GA\v{C}R) project No.~GA22-28797S (D.C., J.C.),
JSPS KAKENHI Grant Numbers 21K13884, 23K03324 (A.H.), 21H01003, 23H03816, 23H03817
(H.S., A.H.), Austrian Federal Ministry of Science, Research and Economy through the Vienna Scientific
Cluster (VSC) Research Center and by the Ministry of Education, Youth and Sports of the Czech Republic through the e-INFRA CZ (ID:90254).
H.S was supported by JSPS KAKENHI Grant Number 21H01041.
H. S. thanks the Supercomputer Center, the Institute for Solid State Physics, and the University of Tokyo for the use of their facilities.
\end{acknowledgments}
%%%%%%%%%%%%%%%%%%%%%%%%%%%%%%%%%%%%%%%%%%%%%%%%%%%%%%%%%%%%%
%%%%%%%%%%%%%%%%%%%%%%%%%%%%%%%%%%%%%%%%%%%%%%%%%%%%%%%%%%%%%

\bibliography{literature1,literature2}

\end{document}